\documentclass[prl,twocolumn,showpacs,amsfonts,amssymb,amsmath,superscriptaddress]{revtex4}

	\usepackage{bm}

 	\usepackage{graphicx}

	\usepackage{hyperref}		
	\usepackage[figure]{hypcap}		

	\hypersetup{
		unicode		= false,	
		pdfborder		= {0 0 0}	
		pdftoolbar		= true,		
		pdfmenubar		= true,		
		pdffitwindow		= false,     	
		pdfstartview		= {FitH},    	
		pdfnewwindow	= true,      	
		colorlinks		= true,		
		linkcolor		= blue,		
		citecolor		= blue,		
		filecolor		= blue,		
		urlcolor			= blue,		
		linktocpage		= true,		
		pdftitle			= {Strained bilayer graphene: Band structure topology and Landau level spectrum},
		pdfauthor		= {M. Mucha-Kruczynski, I.L. Aleiner, V.I. Fal'ko},
		pdfsubject		= {},
		pdfcreator		= {MikTeX},
		pdfproducer		= {Marcin Mucha-Kruczynski},
		pdfkeywords		= {strained bilayer graphene},
}
	\newcommand{\vect}[1]{\boldsymbol{#1}}
	\newcommand{\op}[1]{\hat{\boldsymbol{#1}}}
	\newcommand{\opp}[1]{\tilde{\boldsymbol{#1}}}

	\newcommand{\strain}{\mathcal{A}}
	
\begin{document}

\title{Strained bilayer graphene: Band structure topology and Landau level spectrum}
\author{Marcin Mucha-Kruczy\'{n}ski}
\affiliation{Department of Physics, Lancaster University, Lancaster, LA1~4YB, United Kingdom}
\author{Igor L. Aleiner}
\affiliation{Physics Department, Columbia University, New York, NY 10027, USA}
\author{Vladimir I. Fal'ko}
\affiliation{Department of Physics, Lancaster University, Lancaster, LA1~4YB, United Kingdom}

\begin{abstract}\pdfbookmark[1]{Abstract}{abstract}
We show that topology of the low-energy band structure in bilayer graphene critically depends on mechanical deformations of the crystal which may easily develop in suspended graphene flakes. We describe the Lifshitz transition that takes place in strained bilayers upon splitting the parabollic bands at intermediate energies into several Dirac cones at the energy scale of few meV. Then, we show how this affects the electron Landau level spectra and the quantum Hall effect.
\end{abstract}
\pacs{73.22.Pr,62.20.-x,71.70.Di}
\maketitle

Electrons in bilayer graphene exhibit quite unusual properties: they can be viewed as `massive chiral fermions' 
with parabolic dispersion at intermediate energies and Berry phase $2\pi$ \cite{mccann_prl_2006, novoselov_natphys_2006},
 in contrast to monolayer graphene, where charge carriers are Berry-phase-$\pi$ quasi-particles with linear dispersion 
\cite{novoselov_nature_2005, zhang_nature_2005}. Here, we show that topology of the low-energy band 
structure of electrons in bilayer graphene critically depends on mechanical deformations of the crystal. Strain determines the number of Dirac mini-cones in the low-energy part of the spectrum, below the saddle point in the electron dispersion: 
two with the Berry phases $\pi$ in a strongly strained crystal instead of four (three with Berry phase $\pi$ 
and one with $-\pi$) in an unperturbed crystal \cite{mccann_prl_2006}. These spectral features are tracked down to the evolution of the Landau levels 
for electrons in a magnetic field, and we predict their manifestation in the quantum Hall effect in strained bilayers.

\begin{figure}[b]
\centering
\includegraphics[width=1.0\columnwidth]{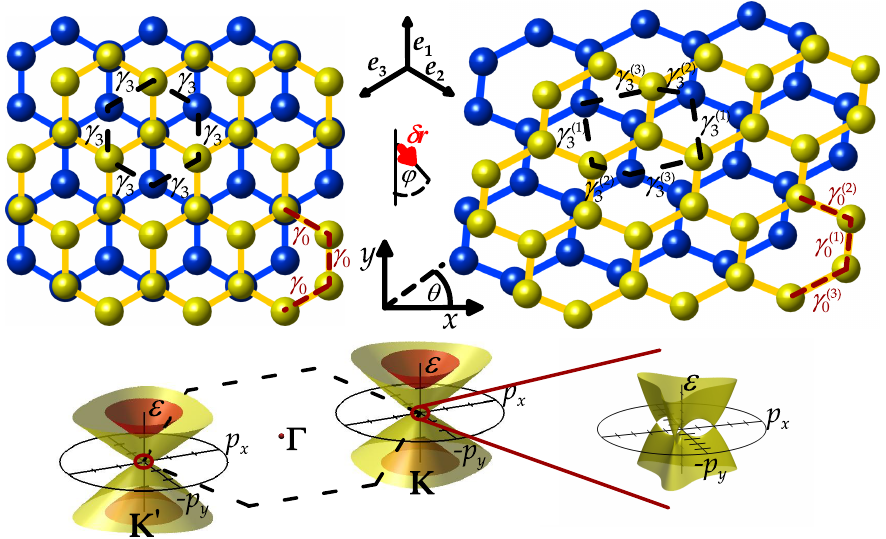}
\caption{Top: Top view of the unperturbed (left) and deformed (right) bilayer graphene lattice. The top and bottom graphene layer is shown in yellow (light) and blue (dark), respectively. Bottom: Electronic band structure of unperturbed bilayer graphene in the vicinity of the Brillouin zone corners {\bf K} and {\bf K'}.}
\label{fig:lattices}
\end{figure}

Bilayer graphene (BLG) consists of two honeycombs of carbons, $A_{1(2)}$, $B_{1(2)}$ in the bottom (top) layer and 
its band structure can be understood using the Slonczewski-Weiss tight-binding model for graphite \cite{slonczewski_physrev_1958_dresselhaus_advphys_1981}.
Within layers, each $A$ site is surrounded by three $B$ sites (and {\it vice versa}), with relative positions given by vectors $\vect{e_{1}}=(0,1)r_{AB}$, 
$\vect{e_{2}}=(\frac{\sqrt{3}}{2},-\frac{1}{2})r_{AB}$ and $\vect{e_{3}}=(-\frac{\sqrt{3}}{2},-\frac{1}{2})r_{AB}$, where $r_{AB}$ is a 
distance between carbon sites (in equilibrium, $r_{AB}\approx 1.4$A), and coupling $\gamma_{0}\sim 3$eV between $\pi$-orbitals of the closest 
carbons. The two layers are arranged according to Bernal stacking shown in Fig. \ref{fig:lattices}: 
sites $A_{2}$ appear on top of $B_{1}$, with the interlayer coupling $\gamma_{1}\sim 0.4$eV, whereas sites $A_{1}$ 
and $B_{2}$ are positioned over the hexagons in the other layer and are coupled by 'skew' 
coupling $\gamma_{3}\sim 0.3$eV. All this gives rise \cite{mccann_prl_2006} to a pair of 'low-energy' bands 
touching each other in the Brillouin zone corners {\bf K} and {\bf K'} which correspond to the electron states 
located on the sites $A_{1}$ and $B_{2}$, and two bands split by the energy $\pm\gamma_{1}$ which correspond to 
electron states located on the sites $A_{2}$ and $B_{1}$ (see Fig. \ref{fig:lattices}).  
The degeneracy of the low-energy bands is the result of cancellation of the contributions 
from the intra- and interlayer hops in directions $\vect{e_{1}}$, $\vect{e_{2}}$ and $\vect{e_{3}}$ 
to the transfer integral for the sublattice Bloch states, which is exact in {\bf K} and {\bf K'}.
 
Perfect crystalline symmetry of graphene can be violated when it is suspended on massive contacts: 
contraction of the latter upon cooling can easily stretch an atomically thin membrane. 
Hence, understanding the influence of deformations on electronic properties of BLG is necessary, 
regarding the growing interest in suspended graphene devices 
\cite{du_natnano_2008, bolotin_ssc_2008, feldman_natphys_2009, weitz_science_2010, martin_prl_2010}.  
Below, we characterise lateral strain in BLG using two principal values, $\delta>\delta'$, 
of the strain tensor $u_{\alpha\beta}=\frac{1}{2}(\partial_{\alpha}u_{\beta}+\partial_{\beta}u_{\alpha})$ 
($\alpha=x$ or $y$ and $\vect{u}=(u_{x},u_{y})$ stand for displacements) and angle $\theta$ between its principal axes and coordinate axes in Fig. \ref{fig:lattices}. In the case of uniaxial strain, the former correspond to the extension along the direction of applied tension ($\delta$) and contraction in the perpendicular direction ($\delta'$). Also, we take into account the discussed-earlier interlayer shear $\vect{\delta\!r}$ \cite{son_arxiv_2010}. 
These deformations make the couplings $\gamma_{0}$ and $\gamma_{3}$ dependent on the direction 
of the hop, which suppresses the above-mentioned exact cancellation in the transfer integral 
for the sublattice Bloch states and results in an additional term in the two-band BLG 
Hamiltonian describing low-energy electrons in the vicinity of the Brillouin zone corners {\bf K} ($\xi=1$) and {\bf K'} ($\xi=-1$): 
\begin{align}\label{eqn:Hamiltonian}
& \!\!\!\op{H} \!=\!  -\frac{1}{2m}\!\left(\begin{array}{cc}
0 & \left(\opp{\pi}^{\dagger}\right)^{2} \\
\opp{\pi}^{2} & 0
\end{array}\!\right) \!+\! \xi v_{3}\!\left(\!\begin{array}{cc}
0 & \opp{\pi} \\
\opp{\pi}^{\dagger} & 0 
\end{array}\!\right) \!+\! \left(\!\begin{array}{cc}
 0 & w \\
w^{*} & 0 
\end{array}\!\right)\!; \\ 
& w = \strain_{3}-\frac{\gamma_{3}}{\gamma_{0}}\strain_{0};\,\,\, \eta_{0/3}=\frac{d \ln \gamma_{0/3}}{d \ln r_{AB}}; \nonumber \\
& \strain_{0}=\frac{3}{4} (\delta-\delta') e^{-2i\theta} \gamma_{0}\eta_{0}\equiv\tilde{\mathcal{A}}_{0}+(\partial_{x}+i\partial_{y})\phi; \nonumber \\ & \strain_{3}=\frac{3}{4} (\delta-\delta') e^{-2i\theta} \gamma_{3}\eta_{3}
-\frac{3}{2} \frac{\delta r}{r_{AB}} e^{i\varphi} \gamma_{3}\eta_{3}, \nonumber \\
& \opp{\pi}=p_{x}+ip_{y}+\tilde{\mathcal{A}}_{0}; \quad \partial_{x}\Im\tilde{\mathcal{A}}_{0}-\partial_{y}\Re\tilde{\mathcal{A}}_{0}\equiv b_{\mathrm{eff}}, \nonumber
\end{align}
where $b_{\mathrm{eff}}$ plays the role of an effective magnetic field \cite{castro_neto_rmp_2009} and $\phi(x,y)$ is determined by the condition that $\Delta\phi=\partial_{x}\Re\strain_{0}+\partial_{y}\Im\strain_{0}$, whereas $\tilde{\mathcal{A}}_{0}=0$ for any homogeneous strain. Also, $m=\frac{2}{9} \hbar^2 \gamma_{1} / r_{AB}^2 \gamma_{0}^2 \approx 0.035m_e$ 
is the effective mass in the parabolic dispersion, $\epsilon\approx\frac{p^2}{2m}$, of electrons at intermediate energies 
$\gamma_1>\epsilon >\mathrm{max}(mv_{3}^{2},|w|)$; and $v_3 = \frac{3}{2} \gamma_{3} r_{AB}/\hbar$. The non-trivial effect of strain in Eq. \eqref{eqn:Hamiltonian} cannot be captured by the theories neglecting the skew coupling $\gamma_{3}$, and  the effect of strain is most significant at low energies, $|\epsilon|\leq\mathrm{max}(mv_{3}^{2}/2,|w|)$.

To derive the effective Hamiltonian \eqref{eqn:Hamiltonian}, we had to take into account that deformations modify 
coupling elements for the intralayer hops $A_{1(2)}-B_{1(2)}$ and interlayer skew hops $A_{1}-B_{2}$ in directions $\vect{e_{n}}$, $n=1,2,3$,
\begin{equation}\label{eqn:hoppings}\begin{split}
& \frac{\gamma_{0}^{(n)}}{\gamma_{0}} =1+\left(\frac{\delta'-\delta}{2}\frac{\vect{e_{n}}}{r_{AB}}\!\cdot\!\vect{l}
+\frac{\delta+\delta'}{2}\right)\eta_{0}, \\  
& \frac{\gamma_{3}^{(n)}}{\gamma_{3}} =1+\left[\frac{\vect{e_{n}}}{r_{AB}}\!\cdot\!\left(\frac{\delta'-\delta}{2}\vect{l}
-\frac{\vect{\delta\!r}}{r_{AB}}\right)+\frac{\delta+\delta'}{2}\right]\eta_{3}.
\end{split}\end{equation}
Here, $\vect{l}=(\sin 2\theta,\cos 2\theta)$. These couplings enter the closest-neighbour tight-binding model for bilayers,
\begin{align*}
\op{H}_{\mathrm{t.b.}} = -\sum_{l=1,2}\sum_{\vect{r_{A_l}}}\sum_{\vect{r_{B_l}}=\vect{r_{A_l}}+\vect{e_{n}}}
\left(\gamma_{0}^{(n)} c_{\vect{r_{A_l}}}^{\dagger} c_{\vect{r_{B_l}}} + \mathrm{H.c.}\right) \\ 
-\sum_{\vect{r_{A_1}}}\sum_{\vect{r_{B_2}}=\vect{r_{A_1}}+\vect{e_{n}}}
\left(\gamma_{3}^{(n)} c_{\vect{r_{A_1}}}^{\dagger} c_{\vect{r_{B_2}}} + \mathrm{H.c.}\right) \\ 
+ \sum_{\vect{r_{A_2}}} \left(\gamma_{1} c_{\vect{r_{A_2}}}^{\dagger} c_{\vect{r_{B_1}}} + \mathrm{H.c.}\right).
\end{align*}
Here, $c^{\dagger}$ ($c$) are creation (annihilation) operators for electrons on the corresponding lattice sites, whereas the vectors $\vect{e_{n}}$ differentiate between three directions of the $A-B$ hops. Note that in Eq. \eqref{eqn:hoppings}, the terms with $(\delta+\delta')$ 
account for `hydrostatic' rescaling of the lattice period and only affect the values of $m$ and $v_3$. 
Also, the direct $A_{2}$-$B_{1}$ interlayer coupling, $\gamma_{1}$, may be changed by shear, 
$\gamma_{1}\rightarrow\gamma_{1}+O\!(r_{AB}^{2})$, without any bearing on the topology of electron bands 
at low energies.

For the electron Bloch states on the sublattice $A_1$, $B_2$, $A_2$ and $B_1$, for the wave vectors  
in the vicinity of the Brillouin zone corners {\bf K} and {\bf K'} (note that strain distorts the hexagonal shape of the Brillouin zone, which we also take into account), 
the four-band Hamiltonian for the electrons has the form
\begin{equation}\label{eqn:4band_Hamiltonian}
\op{H} \!=\!\! \left(\!\!\begin{array}{cccc}
0 & \xi v_{3}\op{\pi} \!+\! \strain_{3} & 0 & \xi v\op{\pi}^{\dagger} \!+\! \strain_{0}^{*} \\
\xi v_{3}\op{\pi}^{\dagger} \!+\! \strain_{3}^{*} & 0 & \xi v\op{\pi} \!+\! \strain_{0} & 0 \\
0 & \xi v\op{\pi}^{\dagger} \!+\! \strain_{0}^{*} & 0 & \gamma_{1} \\
\xi v\op{\pi} \!+\! \strain_{0} & 0 & \gamma_{1} & 0 \\
\end{array}\!\!\right)\!\!, \nonumber
\end{equation}
where $\op{\pi}=p_{x}+ip_{y}$, $v=\frac{3}{2} r_{AB} \gamma_0/\hbar$ is the Dirac velocity in the monolayer. Following the suggestion \cite{castro_neto_rmp_2009} that in monolayers the effect of homogeneous strain appears as 
constant vector potential equivalent to a small shift of the valley centre from the Brillouin zone corners, here,
we employ gauge transformation of the sublattice spinor 
$\psi\rightarrow\psi \exp(-i\xi\phi)$ 
which moves the potenital part $(\partial_{x}+i\partial_{y})\phi$ of $\strain_{0}$ from the anti-diagonal part of the four-band Hamiltonian 
into the diagonal $2\times 2$ block and add (subtract) $\frac{\gamma_{3}}{\gamma_{0}}\tilde{\mathcal{A}}_{0}$ to $\op{\pi}$ $(strain_{3})$ in the diagonal block. 
After this, we use the Schrieffer-Wolff transformation \cite{schrieffer_physrev_1966} 
to project the four-band Hamiltonian onto the pair of low-energy bands \cite{mccann_prl_2006}
describing electron states located predominantly on the sublattices $A_{1}$ and $B_{2}$, and, finally,
arrive at the two-band Hamiltonian in Eq. \eqref{eqn:Hamiltonian}.

\begin{figure*}[t]
\centering
\includegraphics[width=0.85\textwidth]{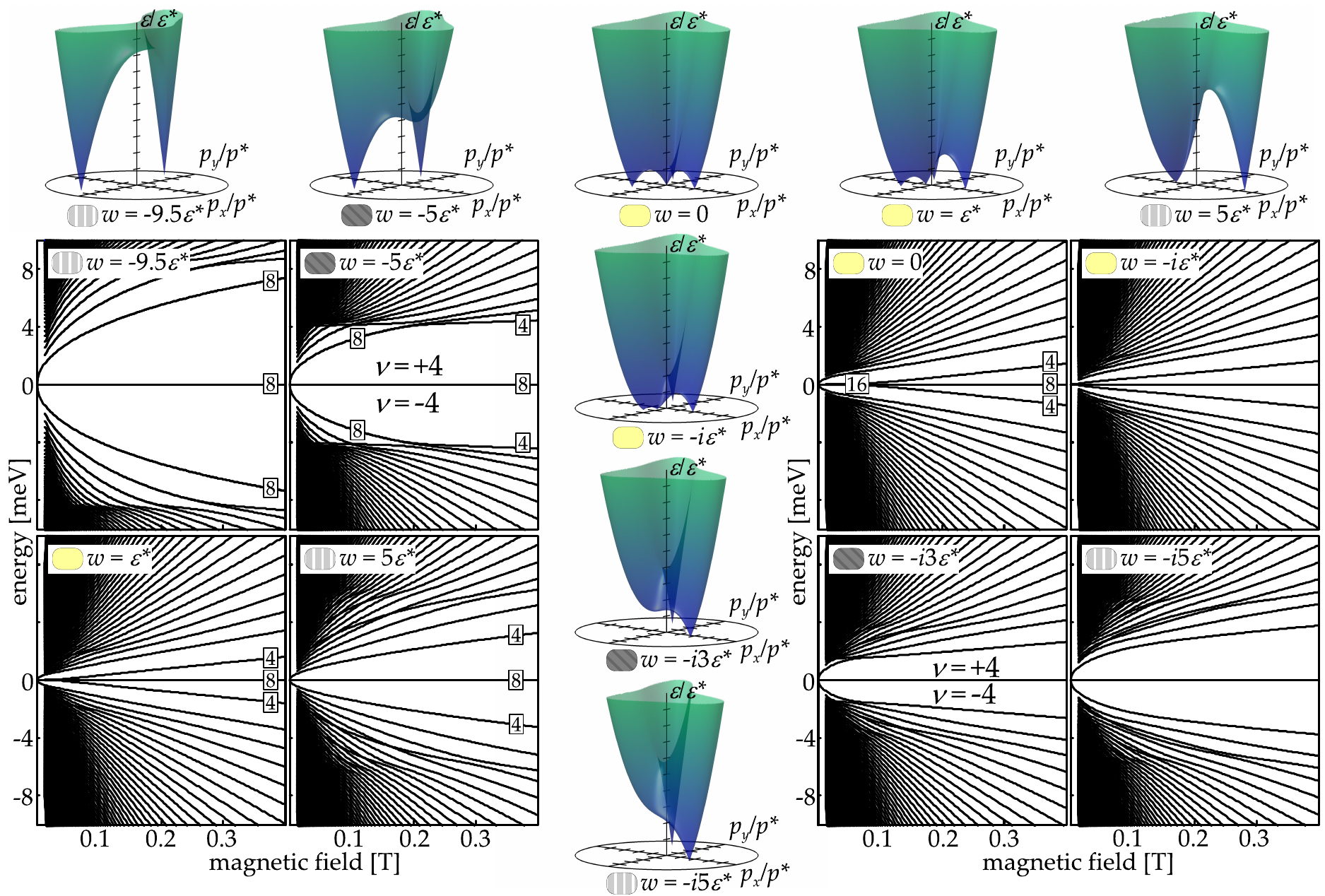}
\caption{Calculated low-energy electronic dispersions in the conduction band of strained BLG and fan plots 
of Landau levels. Dispersion is plotted for the states near the Brillouin zone corners shown in Fig. \ref{fig:lattices} at energies $|\epsilon|<10$meV and for momentum in the units of $p^{*}=mv_{3}$, for several representative points in the $(\Re w,\Im w)$ space, as marked in Fig. \ref{fig:phase_diagram}(a). Spikes at the bottom of dispersion surfaces are the Dirac points characterised by Berry phases $\pm\pi$. For the Landau levels, boxed numbers mark their degeneracy.}
\label{fig:dispersions_and_LLs}
\end{figure*}

To judge the significance of strain, one needs to know the values of lattice parameters $\eta_{0}$ and $\eta_{3}$. Although the value of $\eta_{3}$ is not known, analysis of Raman spectra of monolayers \cite{basko_prb_2009} suggest that $\eta_{0}\approx -3$, and we estimate that 1\% of strain would lead to $|w|\sim 6$meV. 
This can be further enhanced by the electron-electron interaction, which we confirm by  
incorporating the new strain-induced term in the Hamiltoniain \eqref{eqn:Hamiltonian} 
into the earlier-developed renormalization group theory for BLG parameters \cite{lemonik_prb_2010}. 
The calculation, based upon the use of dynamically screened Coulomb interaction 
and the method of $^{1}/_{N}$ expansion ($N=4$ is the number of electron species in BLG: $2\times$spin and $2\times$valley), 
yields in the renormalisation group flow,
\begin{align*}\label{eqn:rg_results}
\partial_{\lambda}w = 0.11w; \, \partial_{\lambda}m^{-1} = -0.02m^{-1}; \, \partial_{\lambda}v_{3}=-0.02v_{3};
\end{align*}
where $\lambda = \ln\frac{\gamma_{1}}{\epsilon}$ and $\epsilon$ is the running energy scale. 
The electron-electron interaction enhances the strain-induced term stronger than other parameters, 
and at energies $\epsilon\sim|w|$, where the influence of strain plays a dominant role in 
determining the electron spectrum, we substitute 
\begin{equation}
|w| \to |w|\exp(0.11\ln\frac{\gamma_{1}}{|w|})\approx |w|^{0.89}\gamma_{1}^{0.11},
\end{equation} 
in the Hamiltonian \eqref{eqn:Hamiltonian}, leading for 1\% strain to an increase from 6meV to the interaction-corrected estimate, $|w|\sim 9$meV.

\begin{figure}[t]
\centering
\includegraphics[width=1.0\columnwidth]{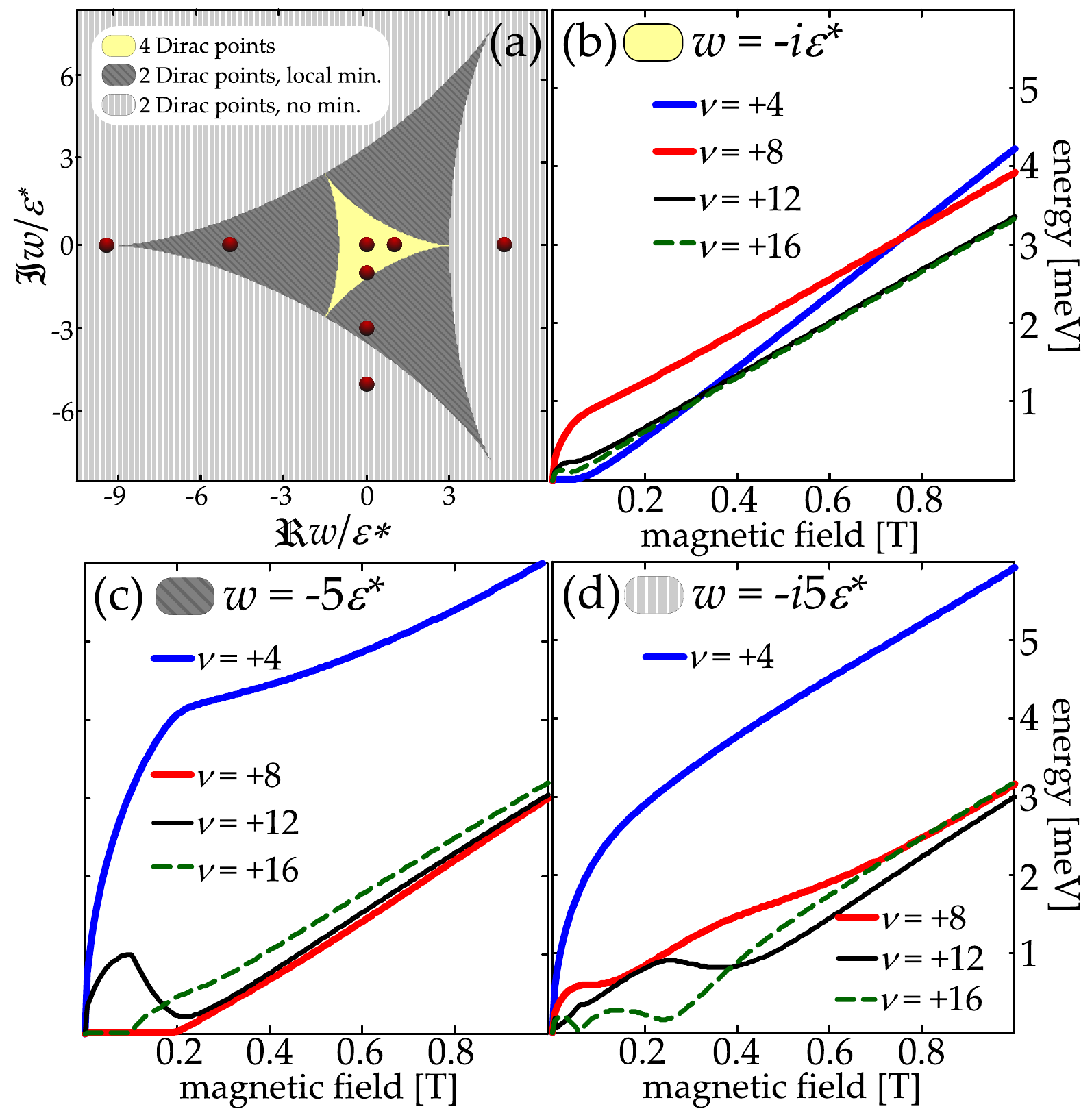}
\caption{(a) Parametric regimes of complex $w$ in Eq. \eqref{eqn:Hamiltonian}, distinguishing between three characteristic topologies of the BLG spectrum in Fig. \ref{fig:dispersions_and_LLs}. 
(b)-(d) Activation energies for the QHE in strained BLG with various integer filling factors. 
For a large enough strain, (c) and (d), filling factor $\nu=\pm 4$ would be the only persistent feature in the low-field QHE.
The appearance of a local minimum in the electron dispersion upon a collision of two Dirac points  
is manifested in (c) by an intermediate saturation of $\epsilon_{\mathrm{act}}(B)$ within the interval $0.2\mathrm{T}<B<0.4\mathrm{T}$.}
\label{fig:phase_diagram}
\end{figure}

The change in topology of the low-energy dispersion for electrons \cite{falko_lectures_windsor} is shown in Fig. \ref{fig:dispersions_and_LLs} 
for several representative values of homogeneous strain chosen from three characteristic regimes distinguished by shading in 
Fig. \ref{fig:phase_diagram}(a). These dispersions are plotted for the conduction band in the valley {\bf K} 
(to be inverted in the momentum space to describe valley {\bf K'} and flipped over 
for the valence band states, $\epsilon\rightarrow-\epsilon$). For $w=0$, electron dispersion undergoes 
Lifshitz transition \cite{lifshitz_jetp_1960} at the energy of the saddle point in the dispersion, $\epsilon^{*}=mv_{3}^{2}/2$: it splits from a single-connected, almost circular line into few disconnected parts, each corresponding to a separate Dirac cone \cite{mccann_prl_2006}. Small strain, $|w|\lesssim\epsilon^{*}$, shifts these Dirac cones across the momentum plane, 
as shown in Fig. \ref{fig:dispersions_and_LLs} for $w=\epsilon^{*}$ and $w=-i\epsilon^{*}$. A stronger strain results in a collision annihilating two Dirac points, 
one with the Berry phase $-\pi$ and another with $+\pi$, which results in a local minimum in the dispersion, as illustrated for $w=-5\epsilon^{*}$ and $w=-3i\epsilon^{*}$.
The other two Dirac points, each with the Berry phase $+\pi$, persist to exist. In Fig. \ref{fig:phase_diagram}(a), the parametric regime where, in addition to a pair of well-separated Dirac cones, the dispersion 
has a local minimum, is marked by dark shading. Finally, much larger strain (light shading in Fig. \ref{fig:phase_diagram}(a)) removes local minimum in the dispersion, 
resulting in even larger separation between the remaining Dirac cones and in a saddle point at $|\epsilon|\approx|w|$, which determines the deformation-dependent 
Lifshitz transition energy in strained BLG. Note that all these spectral changes take place at a relatively low strain, $\sim 1$\%, to compare with the strain of over 20\% \cite{pereira_prb_2009} required to merge Dirac points in monolayer graphene \cite{montambaux_prl_}.

The transformation of electron dispersion by homogeneous strain leads to the modification of the BLG Landau level (LL) spectrum. 
The examples of numerically calculated LLs are shown in Fig. \ref{fig:dispersions_and_LLs} for low magnetic fields, $B<0.4$T. 
Both for small and large strain, the high-magnetic-field end of the LL fan plot, 
$\hbar\omega_{c}\equiv\hbar eB/m\gg\mathrm{max}(\epsilon^{*},|w|)$, is approximately described by 
the sequence $\epsilon\approx\pm\sqrt{n(n-1)}\hbar\omega_{c}$ of four-fold degenerate LLs at 
non-zero energy ($n>2$) and an eight-fold degenerate LL at $\epsilon=0$ ($n=0,1$) \cite{mccann_prl_2006}. 
In non-strained BLG at low fields, such that $\hbar\omega_{c}(B)<mv_{3}^{2}$, 
this transforms into a 16-fold degenerate LL at $\epsilon=0$, so that 
the largest gap in the LL spectrum is between the $\epsilon=0$ and next excited LL, 
suggesting the persistence of filling factor $\nu=\pm 8$ in the quantum Hall effect (QHE) at low magnetic fields. 
After strain causes the annihilation of two out of four Dirac points, 
the $\epsilon=0$ level becomes 8-fold degenerate, and, for strain $|w|\gg\epsilon^{*}$, only filling factors 
$\nu=+4$ and $\nu=-4$ persist in the low-field QHE in BLG: 
the largest energy gap in the LL spectra is between the 8-fold degenerate level at $\epsilon=0$ 
and next excited level, whereas the rest of the spectrum is quite dense. This 8-fold degeneracy is topologically protected and it also appears in a rotationally twisted two-layer stack \cite{de_gail_arxiv_2011}.
Figures \ref{fig:phase_diagram}(b-d) show, for each of the three characteristic regimes of strain in Fig. \ref{fig:phase_diagram}(a), 
how the inter-LL separation (which determines the activation energies, 
$\epsilon_{\mathrm{act}}$, in the QHE) varies from high to low magnetic fields. At the high field end, $\hbar\omega_{c}\gg|w|$,
where the LL spectrum is determined by the quadratic term in equation \eqref{eqn:Hamiltonian}, 
$\epsilon_{\mathrm{act}} \sim \hbar \omega_c \propto B$. For the lowest fields, 
the gap between $\epsilon=0$ and the next LL scales as $\epsilon_{\mathrm{act}}\propto \sqrt{B}$, 
typically for the Dirac-type dispersion emerging upon the Lifshitz transition. 
This behaviour is more pronounced for larger strain. In the regime of intermediate strain corresponding to the 
dark area in Fig. \ref{fig:phase_diagram}(a), 
the activation energy $\epsilon_{\mathrm{act}}$ of the $\nu=4$ state experiences a very unusual intermediate saturation illustrated in Fig. \ref{fig:phase_diagram}(c), indicating that one of the LLs gets stuck in the local dispersion minimum illustrated in Fig. \ref{fig:dispersions_and_LLs}. These results are applicable locally in the case of inhomogeneous strain, using $B_{\mathrm{eff}}=B+\xi b_{\mathrm{eff}}$ \cite{geim_prb_2010} which lifts the valley degeneracy of the LLs, as long as the strain varies smoothly on the scale longer than the effective magnetic length $\lambda=\sqrt{\hbar/eB_{\mathrm{eff}}}$.

To summarize, using homogeneous strain, one can spectacularly change topology of the low-energy electron dispersion, collide and annihilate Berry phase $\pm\pi$ Dirac points near the corners of the Brillouin zone of bilayer graphene crystal. The topological changes in the dispersion of electrons result in the dominance of specifically $\nu=\pm 4$ states in the QHE in BLG at low magnetic fields, with a characteristic behaviour of the activation energy as a function of a magnetic field. The latter features should be viewed in the context of the on-going experimental studies of fundamental properties of bilayers: they give a possibility to 
distinguish the effects of the deformations from spectral changes accompanying the earlier suggested
phase transitions into ferromagnetic \cite{castro_prl_2008} or ferroelectric \cite{zhang_prb_2010, nandkishore_prl_2010} states 
related to the electron-electron interactions,
which are, now, being searched for in suspended BLG devices 
\cite{feldman_natphys_2009, weitz_science_2010, martin_prl_2010}. The latter should differ by opening a gap in the BLG spectrum. However, the ``nematic'' phase of the electron liquid \cite{vafek_prb_2010, lemonik_prb_2010} leads to the same spectral changes as uniaxial strain.

\begin{acknowledgments}
We thank A. Geim, K. S. Novoselov, A. Yacoby and O. Vafek for useful discussions. This work has been funded by EPSRC PhDPlus grant for M.M.-K., EPSRC S\&IA Grant EP/G035954, and EU STREP ConceptGraphene.
\end{acknowledgments}

\end{document}